\documentstyle[prd,aps]{revtex}
\begin{document}
\input epsf
\draft

\def\beq{\begin{equation}}
\def\eeq{\end{equation}}

\renewcommand{\topfraction}{0.8}
\twocolumn[\hsize\textwidth\columnwidth\hsize\csname
@twocolumnfalse\endcsname
\preprint{qqqqqqqqq, hep-ph/-------, qqqqqqqqq, 2000}
\title { \bf Possible Origin of Antimatter Regions in the Baryon 
Dominated Universe} \author{Maxim Yu. Khlopov}
\address{ Center for CosmoParticle Physics "Cosmion", 4 Miusskaya pl., 125047 
Moscow, Russia,\\
Institute for Applied Mathematics, 4 Miusskaya pl., 125047 Moscow, Russia\\
and Moscow Engineering Physics Institute, Kashirskoe shosse 31, 115409 Moscow, 
Russia}
\author{Sergei G. Rubin}
\address{Center for CosmoParticle Physics "Cosmion", 4 Miusskaya pl., 125047 
Moscow, Russia\\
and Moscow Engineering Physics Institute, Kashirskoe shosse 31, 115409 Moscow, 
Russia}
\author {Alexander S. Sakharov}
\address{Labor f\"ur H\"ochenergiephysik, ETH-H\"onggerberg, 
HPK--Geb\"aude, CH--8093 Zurich,\\
 Center for CosmoParticle Physics "Cosmion", 4 Miusskaya pl., 125047 Moscow, 
Russia\\
and Moscow Engineering Physics Institute, Kashirskoe shosse 31, 115409 Moscow, 
Russia}
\date {January 9, 2000}
\maketitle
\begin{abstract}
We discuss the evolution of $U(1)$ symmetric scalar field at the inflation epoch 
with
a pseudo Nambu--Goldstone tilt revealing after the end of exponential expansion 
of the
Universe. The $U(1)$ symmetry is supposed to be associated with baryon charge. 
It 
is shown that quantum fluctuations lead in natural way to baryon 
dominated 
Universe 
with antibaryon excess regions. The range of parameters is
calculated at which the fraction of Universe occupied by antimatter and the
size of antimatter regions satisfy the observational constraints, survive to the 
modern time and lead to 
effects, accessible to experimental search  for antimatter.

\end{abstract}
\pacs{PACS: 98.80.Cq, 11.30Fs, 14.80.Mz.  \hskip 1 cm  (Accepted for
publication in Physical Review D)}
 \vskip2pc]

\section {Introduction}
 The statement that our Universe is baryon asymmetrical as a whole is quite 
firmly 
established observational fact of contemporary 
cosmology. 
Indeed, if large regions of matter and antimatter coexist, then annihilations 
would 
take 
place
at the borders between them. If the typical size of such a domain was
small enough, then the energy released by these annihilations would result in
a diffuse $\gamma$ --ray background, in distortions of the spectrum of the 
cosmic microwave
radiation and light element abundance, neither of which is observed (see for review e.g. \cite{A}). Recent 
analysis of this problem \cite{1} for baryon symmetric Universe 
demonstrates that the size of regions should exceed $1000$ Mpc., being 
comparable 
with the modern cosmological horizon. 
It therefore seems that the Universe is fundamentally matter--antimatter
asymmetric. However the arguments used in \cite{1} do not exclude the
case when the Universe is composed almost entirely of matter with 
relatively small insertions of primordial antimatter. Thus we may expect the 
existence of
macroscopically
large antimatter regions in the Universe, that differs drastically from the case 
of
baryon 
symmetric Universe. We call the region filled with antimatter in the baryon 
dominated 
Universe, as {\it antizillah}. Of course the existence of antizillahs is not 
rigorous
requirement of baryosynthesis, but some modification of baryogenesis
scenarios will result in formation of domains with different sign of 
baryon
charge (see for example \cite{dolg}). The only condition which is necessary 
to
satisfy is the amount of antibaryons within antizillahs must be  small comparing
to 
the total baryon number of the Universe. 

At the first glance it is not difficult to have some amount of antizillahs if 
we simply suppose that in the early Universe when the baryon excess is generated the C--and CP--violation 
have
different sign in different space regions \cite{B}. This 
 may be achieved, for
example,  in models with two different sources of CP--violation, 
explicit and spontaneous \cite{lee} one. However, any spontaneous CP--
violation processes are a result of early phase transition of first or second 
order
 what implies very small size of primordial antizillahs \cite{dolg}. For example if the antizillahs are
formed in the second
order phase transition, their size at the moment of formation is determined by
 $l_i\simeq 1/(\lambda T_c)$, where $T_c$
is so called Ginsburg temperature (the critical temperature at which the phase transition take place) and
$\lambda$ is the selfinteraction  coupling constant of  field which breaks CP
symmetry \cite{lee}. In the result of expansion the modern sizes of domains would reach  
$l_{0}\simeq l_{i}(T_{c}/T_{0})=1/(\lambda T_{0})\simeq 10^{-21}pc/\lambda$, where
$T_{0}$ is the present temperature of the background radiation.

On the other hand it has been revealed \cite{we} that the average
displacement of the
antizillah's boundary  caused by annihilation with surrounding matter is
about $0.5pc$ at the end of 
radiation dominated (RD) epoch. Therefore
any primordial antizillah having initial size up to $0.5 pc$ or more
at the end of RD stage is survived to the contemporary epoch and in the case of 
successive homogeneous expansion has the size 
$\simeq 1kpc$ or more. Any primordial antizillah
with scale less then critical
survival size 
$l_c\simeq 1 kpc$ at contemporary epoch must be eaten up by
annihilation process. Thus it is the serious problem which any model with
thermal phase transition encounters  to create primordial antizillah
with the size exceeding the critical survival size $l_c$ to avoid complete 
annihilation.   

There is an additional problem for  baryosynthesis with
surviving antizillah's sizes. The point is that any phase 
transition is accompanied by formation of
topological defects. If we  blow--up the region with 
different signs
of charge symmetry, we automatically blow--up the scale of respective
topological defect structure. If the structure decays sufficiently late in the 
observable 
part of Universe, the contribution of energy density of such topological
defects could be sufficiently high to contradict with observations. It can be 
easily 
estimated that the structure with the scale corresponding to the survival size 
enters the 
horizon and starts to decay at $T\le 0.1MeV$, i.e. in the period of Big bang 
nucleosynthesis. To remove these
unwanted relics sufficiently early it is necessary to have a 
 mechanism for symmetry restoration. This mechanism implies that the 
baryogenesis is going on within rather narrow time interval 
\cite{dolgsilk,tkachev}. 

In the present paper we have elaborated the issue
for inhomogeneous  baryosynthesis without the difficulties pointed above.
The proposed approach is  based on the mechanism of spontaneous 
baryogenesis \cite{cohen}. This mechanism implies the existence of  
complex scalar field carrying baryonic charge with explicitly broken
$U(1)$ symmetry. The baryon/antibaryon number excess is produced, when the
phase of this additional field moves along the valley of its potential
\cite{cohen,dolgmain}.

It is supposed that the vacuum energy responsible
for inflation is driven by any scalar inflaton field, and additional
complex field coexists with the inflaton. Due to the fact that vacuum
energy during inflational period is
too large, the tilt of  potential is vanished. This implies that the phase of 
the
field behaves as ordinary massless Nambu--Goldstone (NG) boson and the
radius of NG potential is firmly established by the scale of spontaneous
$U(1)$ symmetry breaking. Owing to quantum fluctuations of massless
field at the de Sitter background \cite{alstar,lindebig47} the phase is
varied in different regions of the Universe. 
When the vacuum energy decreases the tilt of potential 
becomes topical, and pseudo NG (PNG) field starts oscillate. As the field rolls 
down in one 
direction during the first oscillation, it preferentially creates baryons over
antibaryons, while the opposite is true as it rolls down in the opposite direction.
Thus to have globally baryon dominated Universe one must have the 
phase sited in the point, corresponding to the positive baryon excess 
generation, 
just 
at the 
beginning of inflation (when the size of the modern Universe crosses the 
horizon). Then subsequent quantum fluctuations can move the phase to the
appropriate position causing the antibaryon excess production. If it takes place 
not 
too late after the inflation begins, the size of antizillah may exceed the 
critical 
surviving size $l_c$.

The main idea of proposed issue is based on the existence of quantum
fluctuations along the effectively massless angular direction of
baryonic charged scalar field. Thus, more general, the considered issue of
generation of antizillahs is applicable practically to all mechanisms of
baryogenesis where the number density and sign of baryon asymmetry depend
on the angular component of complex scalar field. The advantage of the
mechanism of spontaneous baryogenesis  \cite{cohen} considered here is
the quite simple unambiguous inflation dynamics of scalar field generated
baryon charge. This fact allows to establish quantitatively definite
relationship between the effects of inflation and generation of baryon
(antibaryon) excess in inhomogeneous baryogenesis. However, this
relationship may be too rigid for the realistic model of antimatter domain
formation compatible with the whole set of astrophysical constraints. The
consistent picture may need more sophisticated scenarios. The principal
possibility for such scenario can be considered on the base of Affleck
Dine (AD) \cite{afflekdine} baryogenesis mechanism that still
receives a lot of attention \cite{afflekdine,lisa,b1,b2,b3,mr}. 

AD baryogenesis also involves the cosmological evolution of
effective scalar field, 
which carries baryonic charge, being composed of
supersymmetric partners of
electrically neutral quark and lepton combinations. The important
feature of supersymmetric extensions of standard model  is the existence
of  "flat directions" in field space, on which the scalar potential
vanishes \cite{nil,afflekdine,lisa}. We will refer for the definiteness to
the flat directions of minimal standard supersymmetric model (MSSM)
\cite{lisa,mr1}. Thus, if the some component of scalar field lies
along a flat direction, this component can be considered as a free massless
complex scalar so called AD field \cite{afflekdine,lisa}. At the level of
renormalizable terms, "flat directions" are generic, but supersymmetry
breaking and nonrenormalizable operators lift the "flat directions" and
sets the scale for their potential. During the inflational period the AD
field develops non--zero vacuum expectation value and subsequently when
the Hubble rate becomes of the order of the curvature of AD potential, the
condensate starts to oscillate around its present minimum. Baryon
asymmetry can be induced in such condensate only if there exists phase
shift between real and imaginary parts of the AD field. Such shift and
consequently B and CP violation is provided by A--term in the potential
which parameterizes MSSM "flat direction" \cite{afflekdine,lisa}. The
resulting
sign and number density of baryon asymmetry depends on the magnitude of
initial phase of AD field and on phase shift created by A-term at the
relaxation period \cite{afflekdine,lisa,b1}. Therefore the de--Sitter
fluctuations can generate antizillahs in the baryon asymmetric Universe in
the similar way to the spontaneous baryogenesis if the angular direction
of AD field is characterized the mass that is much smaller that the Hubble
constant $H$ during inflation. It takes place if there are no of order $H$
corrections to the A--term \cite{mr}. 

The early dynamics of AD field is quite complicated \cite{b3} owing
to the non--trivial background energy density driving inflation in
MSSM. Moreover AD potential can get corrections from the vacuum energy that
removes its minimum from the original one \cite{lisa,b1,b3,mr}. In general
there are two types of inflation in MSSM, D--term or F--term inflation
(see for review \cite{riot}), depending on the type of vacuum contributing
the energy density during de-Sitter stage. In the case of D--term
inflation AD fields and inflaton slow roll coherently \cite{b3} (in the
absence of order $H^2$ corrections to the mass squared term of AD
potential). It implies that the radius of effectively massless angular AD
direction is determined by the immediate value of inflaton field.
For the case of F--term inflation the AD scalar will get an
order $H^2$ negative mass squared term \cite{lisa,b1,b3,mr} causing the
minimum of AD potential. The AD field is closed to the minimum during
the F--inflation stage \cite{b3} and this minimum determines the radius
of circle valley of effectively massless angular direction. 

The conclusion from this explicit example based on the MSSM is following.
For any complicated inflation dynamics of baryon charged field it
is possible to simulate appropriate massless direction that behaves
similar to the circle valley of NG potential. This fact makes the proposed
issue for generation of antizillahs viable not only for spontaneous
baryogenesis mechanism, but for the all mechanisms dealing with
effectively massless angular directions during inflation \cite{other}.    

The paper is organized as to following. In section II we discuss 
the 
quantum behavior of nondominant $U(1)$ symmetric scalar field at the 
inflation period. We estimate the amplitude and space scale of fluctuations of 
the phase for this field without PNG tilt. The size distribution 
of these fluctuations determines the size distribution of antizillahs. 
The section III contains calculations of 
baryon/antibaryon net excess production at the relaxation of  phase when the
tilt of Mexican hat potential becomes topical. We summarize our conclusions and 
discuss some problems of the considered scenarios in section IV.

\section{\label{PERT} Phase Distribution for NG Field at The Inflation Period}
We start our consideration with the discussion of evolution of $U(1)$ symmetric  
scalar field which coexists with inflaton at the inflation epoch. The  quantum 
fluctuations of such field during the inflation stage cause the perturbations 
for the phase marking the Nambu--Goldstone vacuum. In our model this phase
determines the sign and value of baryon excess, so the size distribution
of domains containing the appropriate phase values, caused by that 
fluctuations, coincide with the size distribution of antizillahs.  

Thus to estimate the number density of antimatter regions with sizes exceeding 
the
critical survival size $l_c$ in the baryogenesis model under consideration we 
have 
to 
deal with long -- wave quantum fluctuations of the NG boson field at the 
inflation 
period. 
Various aspects of this question have been examined in the numerous papers 
\cite{kofm,lindekofm,spokyok,lindelyth,lindeax,lindebook,lyth,lythstewart,lindebig} 
in the connection with cosmology of invisible axion. Also the de--Sitter
quantum fluctuations have been analyzed in the framework of AD
baryogenesis \cite{b2,b3}.

 The effective potential of the complex field is taken in the usual form

\beq
\label{3phase}
V(\chi )=-m_{\chi}^2\chi^*\chi+\lambda_{\chi}(\chi^*\chi )^2+V_0,
\eeq
where the field $\chi$ can be represented in the form

\beq
\label{phase}
\chi =\frac{f}{\sqrt{2}}\exp{\left(\frac{i\alpha}{f}\right)}
\eeq

The $U(1)$ symmetry breaking implies that the radial component of the field 
$\chi$ 
acquires a 
nonvanishing classical part, $f=m_{\chi}/\sqrt{\lambda_{\chi}}$ and field 
$\alpha$ in eq. 
(\ref{phase}) becomes a massless NG scalar field with a vanishing 
effective potential,
$V(\alpha )=0$. In this case, $\chi$  has the familiar Mexican--hat potential, 
and the degenerated vacua correspond to the circle of radius $f$. Throughout 
present paper we deal with dimensionless 
angular field 
$\theta =\alpha /f$.

We concern here the possibility to 
store appropriate phase value in the domain with the size exceeding the critical 
survival 
size. Such value of phase plays the role of starting point for clockwise 
movement, which is going to generate antibaryon excess when the
tilt of potential breaking $U(1)$ explicitly, will turn to be topical. 

We assume that the Hubble constant varies slowly during inflation.
Also we use well established behavior of quantum fluctuations on the de Sitter
background  \cite{lindebook}. It implies that vacuum
fluctuations of every scalar field grow  exponentially in the inflating
Universe. When the wavelength of a particular fluctuation becomes greater
than $H^{-1}$ the average amplitude of this fluctuation freezes out at some 
nonzero
value because of the large friction term in the equation of motion  of the
scalar field, whereas its wavelength grows exponentially. In the
other  words such a frozen fluctuation is equivalent to the appearance of
classical field  that does not vanish after averaging over macroscopic space
intervals. Because the  vacuum must contain fluctuations of every wavelengths,
inflation leads to the  creation of more and more new regions containing the
classical field of different amplitudes with  scale greater than $H^{-1}$. The 
averaged amplitude of such NG field fluctuations generated during each time 
interval $H^{-1}$ is given by  \cite{alstar}
\beq  
\label{b11}  
\delta\alpha =\frac{H}{2\pi}
\eeq
During such time interval the universe expands by a factor of $e$.  
Since the NG field is massless during inflation 
period (the PNG tilt is vanish yet), one can see that the amplitude of each 
frozen fluctuation is not 
changed in 
time at all and  the phases of each wave are random. Thus the quantum evolution 
of 
NG field looks  like one--dimensional Brownian motion \cite{lindebook,lindebig} 
along the circle  valley corresponding to the bottom of NG potential. This 
statement implies that the  values of the phase $\theta$ in different regions
become 
different, and  the corresponding variance grows as \cite{lindebig47} 
\beq 
\label{b12} 
\langle (\delta\theta )^2\rangle =\frac{H^3t}{4\pi^2f^2} 
\eeq 
that means that dispersion grows as  
$\sqrt{\langle (\delta\theta )^2\rangle}=\frac{H}{2\pi f}\sqrt{N}$, where N is 
the  number of e--folds. In the other words the phase $\theta$ makes quantum 
step 
with the  
scale $\frac{H}{2\pi f}$ at each e--fold, and the total number of steps during 
some 
time  interval $\Delta t$ is given by $N=H\Delta t$.

Let us consider the scale $k^{-1}=H_0^{-1}=3000h^{-1}Mpc$ which is the 
biggest  cosmological scale of interest. We suppose that Universe is baryon 
asymmetric in this  
scale which leaves the horizon at definite e--fold $N=N_{max}$. On the other 
side 
this 
scale is the  
one entering the horizon now, namely $a_{max}H_{max}=a_0H_0$ where 
the  
subscript $0$ indicates the contemporary epoch. This implies that: 
\beq 
\label{b13} 
N_{max}=\ln{\frac{a_{end}H_{end}}{a_0H_0}}-
\ln{\frac{H_{end}}{H_{max}}} 
\eeq 
the subscript $end$ denotes the epoch at the end of inflation. The 
slow-roll  
paradigm tells us that  the last term of (\ref{b13}) is usually $\le 1$. The 
first 
term  
depends on the evolution of scale factor $a$ between the end of slow-roll 
inflation  
and the present epoch. Assuming that inflation ends by short matter dominated 
period, which is followed by RD stage lasting until the present 
matter  
dominated era begins, one has \cite{LythLiddlePhysRep} 
\beq
\label{b14} 
N_{max}=62-\ln{\frac{10^{16}GeV}{\sqrt{H_{end}M_p}}}- 
\frac{1}{3}\ln\frac{\sqrt{H_{end}M_p}}{\rho_{reh}^{1/4}}, 
\eeq 
where $\rho_{reh}^{1/4}$ is the reheating temperature when the RD stage is 
established.  
With $H_{end}\simeq 10^{13}GeV$ and instant reheating this gives  
$N_{max}\approx 62$, the largest possible value. However, if one has to
invoke supersymmetry to prevent the flatness of the inflation potential,
for example like as in the case of AD baryogenesis, the $\rho_{reh}^{1/4}$
should not exceed then $10^{10}GeV$ to avoid too many gravitino
overproduction \cite{KhlopovLinde}, and one have $N_{max}=58$, perhaps the
biggest  reasonable   value. Through the paper we will use $N_{max}=60$.
The smallest  cosmological scale of antizillah that is survived after
annihilation  is  
$k^{-1}_{c}=l_{c}\approx 8h^2kpc$ \cite{we}. It is $9$ order of magnitude 
smaller 
then $H^{-1}_0$, 
that  
corresponds to 
\beq 
\label{b15} 
N_c\approx N_{max}-13-3\ln{h}\approx 45 
\eeq 
Thus the $l_c$ should left horizon at 45--folds before the end of inflation. 

\begin{figure}[t]
\centering
\leavevmode\epsfysize=5.1cm \epsfbox{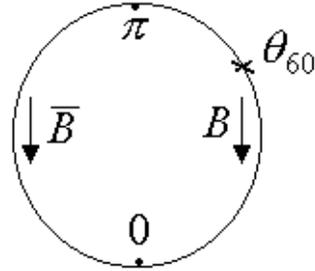}\\
\
\caption[fig2]{\label{fig1} Baryosynthesis in the spontaneous baryogenesis mechanism. 
The sign of baryon asymmetry depends on the starting point of phase 
oscillations.}
\end{figure} 

 Let us assume that the phase value $\theta =0$ corresponds to South Pole
of  NG field  
circle valley, and $\theta =\pi$ corresponds to the opposite pole. The positive 
gradient
of phase in this picture is routed as anticlockwise direction, and the dish
of PNG  potential  would locate at the South Pole of circle (see
fig.\ref{fig1}). It  will be 
shown 
below (see 
section III) that the antibaryon production  corresponds to the regions that 
would 
contain phase values caused anticlockwise rolling  
of PNG field $\alpha$ during the first half period of oscillation.  
If the field $\alpha$ rolls clockwise towards the dish of tilted potential 
just after the start of first oscillation then baryon production will take 
place.  

Now we are in the position to estimate the fraction of the Universe containing
antizillahs. To ensure that the 
Universe would be baryon asymmetric as a whole it is necessary to suppose that 
the phase average value $\theta =\theta_{60}$ within biggest cosmological scale 
of interest emerging at the $N_{max}=60$ e--folds before the end of inflation is  
located in the range $[0,\pi ]$. The $\theta_{60}$ is the starting 
point for Brownian motion of the phase value along the circle valley during 
inflation. As it has been 
mentioned  above, the phase makes Brownian step $\delta\theta  =\frac{H}{2\pi f}$ 
at each e--fold. Because the typical wavelength of the 
fluctuation $\delta\theta$  
generated during such timescale is equal to $H^{-1}$, the whole domain 
$H^{-1}$,  
containing $\theta_{60}$, after one e--fold effectively becomes divided into 
$e^3$  separate, causal disconnected domains of radius $H^{-1}$. Each domain
contains almost homogeneous  phase value 
$\theta_{60-1}=\theta_{60}\pm\delta\theta$. In half of these 
domains the phase evolves towards $\pi$ (the North Pole) and in the other 
domains 
it moves  towards zero (the South Pole). To have antizillah with
appropriate  sizes to avoid full annihilation one 
should 
require that the phase value crosses 
$\pi$ or zero  not later then after $15$ steps. Only in this case the 
antizillahs 
would have the sizes larger than $l_c$ and are conserved up to the modern era. 
This 
means that one of the two following inequality  must be satisfied 
\beq
\label{b16} 
\pi -\frac{15H}{2\pi f}\le\theta_{60}\le\frac{15H}{2\pi f}
\eeq 
Consider initially the case of exact equalities in expression (\ref{b16})  when 
the 
main part of antimatter is contained in the antizillahs of size $l_c$. The 
number of 
domains containing the equal values of phase at the $45$ e--folds before the   
end 
of inflation is given by the following expression
\beq
\label{b17a}
n_{45}\approx (e^3/2)^{15}\approx 10^{15}. 
\eeq
Then the probability that every domain of size $l_c$
would  not be separated into $e^3$  domains with size one order
of magnitude less then $l_c$ at the next e--fold is given by
$P_s\approx (1/2)^{e^3}\approx 10^{-6}$. Thus the number of domains 
serving
as the prototypes for antizillahs of size $l_c$ looks 
like  
\beq 
\label{b18} 
\bar n=n_{45}P_s\approx 10^9 
\eeq 
There are about $10^{11}$ galaxies in the Universe. Thus, according to 
such simple consideration, we reveal that $1\%$ of volume boxes
corresponding to each galaxy contains the region of  size  $l_c$
filled with  
antimatter of highest possible antibaryonic density if the $\theta_{60}$ 
coincides 
with left side of inequality (\ref{b16}) or lowest one in the case  if the 
opposite equality is held. 

We are able also to find the size distribution for antizillahs. For this purpose 
it is necessary to study the inhomogeneities 
of phase induced by (\ref{b11}). It has been well established that for any given 
scale 
$l=k^{-1}$ large scale component of the phase value $\theta$ is distributed in 
accordance with Gauss's law \cite{alstar,lindebig47,lindebook,lindebig}. The 
quantity which will be especially interesting for us is the dispersion
(\ref{b12}) for quantum fluctuations of phase with moments from $k=H^{-1}$ to 
$k_{min}=l^{-1}_{max}$ (where the $l_{max}$ is the biggest cosmological 
scale that corresponds to $60$ e--folds). This quantity can be expressed in the 
following manner 
\beq
\label{b19}
\sigma^2_l=\frac{H^2}{4\pi^2}\int\limits_{k_{min}}^kd\ln{k}=\nonumber\\
\frac{H^2}{4\pi^2}\ln{\frac{l_{max}}{l}=\frac{H^2}{4\pi^2f^2}(60-N_l)},
\eeq
where $N_l$ is the number of e--folds which relates the biggest cosmological 
scale to the given scale $l$.  
This means that the distribution of phase has the Gaussian form
\beq
\label{b21}
P(\theta_l ,l)=\frac{1}{\sqrt{2\pi}\sigma_l}
\exp{\left\{ -\frac{(\theta_{60}-\theta_l)^2}{2\sigma^2_l}\right\}}
\eeq

Suppose that at e--fold $N_t$ 
before the end of inflation the volume $V(\bar\theta ,N_t)$ has been filled with 
phase value $\bar\theta$. Then at the e--fold $N_{t+\Delta t}=N_t-\Delta N$ the 
volume filled with phase $\bar\theta$ will follow iterative 
expression
\begin{eqnarray}\label{iteration}
V(\bar\theta , N_{t+\Delta t})=e^3V(\bar\theta ,N_t)+\nonumber\\
+(V_U(N_t)-e^3V(\bar\theta ,N_t)P(\bar\theta , N_{t+\Delta t})h.
\end{eqnarray}
Here the $V_U(N_t)\approx e^{3N_t}H^{-3}$ is the volume of the Universe at $N_t$
e--fold. Expression (\ref{iteration}) makes it possible to calculate the size 
distributions 
of domains 
filled with appropriate value of phase numerically. In order to illustrate 
quantitatively the number distribution of domains, we 
present here the numerical results for definite values of $\theta_{60}$, 
$\bar\theta$ and $h=\frac{H}{2\pi f}$. The table contains the results concerning 
to number of domains with average phase $\bar\theta$ at e-fold number $N$, 
\begin{table}
\caption{The sample of distribution of proto--antizillahs by sizes and numbers 
of  
e--folds at $ \theta _{60}=\frac{\pi }{6}$;
$\overline\theta  =-0$; $h=0.026$}
\begin{tabular}{lcc}
$N$&$N_{antizillahs}$&$L_{antizillah}h$ \\ \tableline
$59$&$0$&$1103$Mpc\\
$55$&$5.005 \, \times \, 10^{-14}$&$37.7$Mpc\\
$54$&$7.91 \, \times \, 10^{-10}$&$13.9$Mpc\\
$52$&$1.291 \, \times \, 10^{-3}$&$1.9$Mpc\\
$51$&$0.499$&$630$kpc\\
$50$&$74.099$&$255$kpc\\
$49$&$8.966 \, \times \, 10^{3}$&$94$kpc\\
$48$&$8.012 \, \times \, 10^{5}$&$35$kpc\\
$47$&$5.672 \, \times \, 10^{7}$&$12$kpc\\
$46$&$3.345 \, \times \, 10^{9}$&$4.7$kpc\\
$45$&$1.705 \, \times \, 10^{11}$&$1.7$kpc\\
\end{tabular}
\end{table}

The fraction of the Universe filled with phase $\bar\theta$ appears to be equal 
to $7.694\times 10^{-9}$.
Thus we see that the distribution of domains with size is very abrupt and should 
be 
peaked at smallest value of size. Adjusting the 
free 
parameters $\theta_{60}$ and $h$ we are able to achieve the situation that 
volume box corresponding to each galaxy contains ($1\div 10$) regions
with appropriate  phase $\bar\theta$. The sizes of such regions are larger
or equal to critical surviving  size.  In spite of the sufficiently large
total number of antizillahs only the  small 
part of our Universe will be occupied by antizillahs (see the last line in the 
presented 
table). 

The nontrivial question on the actual forms of astrophysical objects
antizillahs can have in the modern Universe needs spacial analysis, which,
in general, strongly depends on the assumed form of the nonbaryonic dark
matter, dominating in the period of galaxy formation. However, based on
the early analysis \cite{we,khlop,ams} the two extreme cases can be
specified, when the evolution of antizillahs is not strongly influenced by
the dark matter content. In the first case, the antibaryon density within
the antizillah is by an order of magnitude higher than the average baryon
density, so that the over-density inside this region can exceed the dark
matter density and rapid evolution of such an antizillah with the size
exceeding the surviving scale can provide formation of compact antimatter
stellar system (globular cluster (see for review \cite{glob})) which
can survive in galaxy \cite{khlop,ams}. The other extreme case is
antizillah with extremely low internal antibaryon density
$\Omega_{\bar B}<10^{-5}$. Then the diffused antiworld is realized, when
no compact antimatter objects are formed and antizillahs evolve
into low density antiproton-positron plasma regions in voids outside the
galaxies \cite{we,khlop}.

\section {\label{BARYONS} Spontaneous Baryogenesis Mechanism }
The following element of our scenario of inhomogeneous baryogenesis should 
contain a 
conversion of the phase $\theta$ into baryon/antibaryon excess. We
consider the ansatz of spontaneous baryogenesis mechanism \cite{cohen}. The
basic feature of this mechanism is that the 
sign 
of baryon charge created by relaxation of energy of PNG field 
critically depends on the direction that the phase is rotated on the bottom of 
Mexican heat potential. It provides us to convert the domains 
containing the appropriate phase value, caused by fluctuations, to 
the antizillahs at the period when the NG potential gets the tilt. 

The one of reasonable issue to the spontaneous baryogenesis \cite{cohen}
has been considered  in the work \cite{dolgmain}. Let us briefly discuss
it.  It was assumed that in the early Universe a complex scalar field 
$\chi$ coexists with  inflaton $\phi$ responsible for inflation. This field 
$\chi$ has non vanishing baryon number. The possible 
interaction of $\chi$  that violates lepton number can be described by following
Lagrangian density (see e.g. \cite{dolgmain})
\begin{eqnarray}
\label{1}
L=-\partial_{\mu}\chi^*\partial^{\mu}\chi -V(\chi )+i\bar 
Q\gamma^{\mu}\partial_{\mu}Q +
i\bar L\gamma^{\mu}\partial_{\mu}L - \nonumber\\ -m_Q\bar QQ-
 m_L\bar LL +
(g\chi\bar QL+h.c.)
\end{eqnarray}
The fields $Q$ and $L$ could represent heavy quark and lepton, 
coupled to the 
ordinary quark and lepton matter fields. Since fields $\chi$ and $Q$ possess 
baryon number while the 
field $L$
does not, the 
couplings in the (\ref{1}) violate lepton number \cite{dolgmain}. The 
$U(1)$ symmetry  that 
corresponds to baryon number is expressed by following transformations
\beq
\label{2}
\chi\to\exp{(i\beta )}\chi ,\qquad Q\to\exp{(i\beta )}Q,\qquad L\to L
\eeq
The effective Lagrangian density for $\theta$, $Q$ and 
$L$ eventually has the 
following form after symmetry breaking \cite{dolgmain}
\begin{eqnarray}
\label{5}
L=-\frac{f^2}{2}\partial_{\mu}\theta\partial^{\mu}\theta
+i\bar Q\gamma^{\mu}\partial_{\mu}Q +
i\bar L\gamma^{\mu}\partial_{\mu}L - \nonumber\\ -m_Q\bar QQ 
-m_L\bar LL +
(\frac{g}{\sqrt{2}}f\bar QL+h.c.)+\partial_{\mu}\theta\bar 
Q\gamma^{\mu}Q
\end{eqnarray}
At the energy scale $\Lambda <<f$, the symmetry (\ref{2}) is explicitly
broken 
 and the Mexican--hat circle gets a little pseudo NG tilt described by the 
potential
\beq
\label{tilt}
V(\alpha )=\Lambda^4(1-\cos{\theta})
\eeq
This potential, of high $2\Lambda^4$, has a unique minimum at 
$\theta =0$. Of course, in the most cases, the potential (\ref{tilt}) 
is 
the lowest--order approximation to a more complicated expressions emerged 
from particle physics models (see e.g. \cite{freese} and Refs. therein). 

The important parameter for spontaneous 
baryogenesis is the curvature of (\ref{tilt}) in the vicinity of its minimum, 
which is determined by the mass of PNG field
\beq
\label{mass}
m^2_{\theta}=\frac{\Lambda^4}{f^2}
\eeq
As it was mentioned above the field $\chi$ is an additional field with
nondominant energy density contribution to the Habble constant deriving by
de Sitter stage. Suppose that the tilt was formed during inflation. Then the
order of magnitude estimation for fluctuations induced by large-- scale
inhomogeneity of oscillations of the field $\chi$ gives $\frac{\delta
T}{T}=\frac{1}{3}\frac{\delta \rho}{\rho}~(\Lambda /T)^4$. Thus, for
$T=H/2\pi$ and reasonable value $\Lambda\simeq 10^{-5}H$ (see the end
of this section) the thermal electromagnetic background
fluctuations are within the observational limits. 

Also we assume that the field $\theta$ behaves as massless NG  
field during inflation implying that  the condition  
\beq
\label{cond}
m_{\theta}<<H
\eeq
is valid, where the $H$ is the Hubble constant during the inflation. After the 
end of inflation condition (\ref{cond}) is violated and the oscillations of 
field $\theta$ around the minimum of  potential (\ref{tilt}) are started. The 
energy density $\rho_{\theta}\simeq\theta^2_im_{\theta}^2f^2$
of the PNG field which has been created by quantum fluctuations of
$\theta$  during the inflation converts to baryons and antibaryons
\cite{cohen,dolgmain}. The  sign of baryon charge depends on the initial
value of phase from which the  oscillations are started.  

Let us estimate the number of baryons and antibaryons produced by classical 
oscillations of field $\theta$ with an arbitrary initial phase $\theta_i$. 
The 
appropriate expression for the density of produced baryons (antibaryons)
$n_{B(\overline{B})}$ is represented in \cite{dolgmain} 
\beq 
\label{n1} 
n_{B(\overline{B})}=\frac{g^{2}}{\pi ^{2}}\int\limits_{m_{Q}+m_{L}}^{\infty 
}\omega 
d\omega \left| \int_{-\infty }^{\infty }dt\chi (t)e^{\pm 2i\omega t}\right|^{2},  
\eeq 
that is valid if $\chi (t\rightarrow -\infty )=\chi (t\rightarrow +\infty )=0.$ 
General case can be obtained in the limits 
$\chi (t\rightarrow -\infty )\neq 0;\chi (t\rightarrow +\infty )=0$\ without loss of generality. After 
integration by part expression (\ref{n1}) has the form 
\beq
\label{bar}
N_{B(\bar B)}=\frac{g^{2}}{4\pi ^{2}}\Omega _{\theta _{i}}\int
 \limits_{m_{Q}+m_{L}}^{\infty}d\omega 
\left| \int\limits_{-\infty }^{\infty }d\tau \dot{\chi}(\tau )e^{\pm 
2i\omega \tau }\right| ^{2}, 
\eeq 
where the $\Omega_{\theta_i}$ is the volume containing the phase value 
$\theta_i$.
Here the surface terms appear to be zero at $t=\infty$ due to asymptotic of 
field $\chi $ and at $t=-\infty $ due to Feynman radiation conditions.

For our estimations it is enough to accept that the phase changes 
as
\beq
\label{osc}
\theta (t)\approx\theta_i(1-m_{\theta}t)
\eeq 
during first oscillation. We also
put $m_Q=m_L=0$
that is reasonable for estimations. Substituting (\ref{osc}) and (\ref{phase}) 
into 
(\ref{bar}) we come to
\beq
\label{bar1}
N_{B(\bar B)}\approx\frac{g^2f^2m_{\theta}}{8\pi^2}
\Omega_{\theta_i}\theta^2_i\int
\limits_{\mp\frac{\theta_i}{2}}^{\infty}d
\tilde\omega\frac{\sin^2{\tilde\omega}}{\tilde\omega^2},
\eeq
where the sign in the lower limit of integral corresponds to baryon or 
antibaryon net excess generation respectively. The reasonability of our 
approximation follows from comparison of (\ref{bar1}) at small 
$\theta_i<<1$ 
\beq
\label{barapprox}
N_B-N_{\bar B}=\frac{g^2f^2m_{\theta}}{8\pi^2}
\Omega_{\theta_i}\theta^3_i
\eeq
with the result of \cite{dolgmain}.  

Using for spatially homogeneous field $\chi=\frac{f}{\sqrt{2}}e^{i\theta }$ the 
expression for baryon charge 
\beq
\label{rub1}
Q=i(\chi ^{\ast }d\chi /dt-d\chi ^{\ast }/dt\chi )=-fd\theta /dt, 
\eeq 
one can easily conclude that $Q>0$ if $\theta >0$ during classical movement 
of phase $\theta $ to zero. Thus the anticlockwise rotation gives rise to
antibaryon  excess while the clockwise rotation to the baryon excess
one.

During reheating, the inflaton energy converts into the radiation. It is assumed 
that 
reheating 
takes place when the Mexican--hat potential is not sensitive to the PNG tilt 
yet. This 
implies that the total decay width of inflaton $\Gamma_{tot}$ into light degrees 
of
freedom exceeds the mass $m_ {\theta}$. In the other words the reheating is 
going 
on under the condition (\ref{cond}). The relaxation of $\theta$ field starts 
when 
$H\approx m_{\theta}$ and converts to the baryons or antibaryons. Baryonic 
charge 
is converted inside a comoving volume after reheating owing to very effective 
decay 
during the cosmological time. This means that the baryon--to--entropy ratio in
$n_{B(\bar B)}/s=Const$ in the course of expansion. The entropy density after 
thermalization is given by
\beq
\label{entropy}
s=\frac{2\pi^2}{45}g_*T^3
\eeq
where $g_*$ is the total effective massless degrees of freedom. Here we 
concern with the temperature above the electroweak symmetry breaking scale. At 
this temperature all the degrees of freedom of the standard model are in 
equilibrium 
and $g_*$ is at least equal to $106.75$. The temperature is connected with 
expansion rate as follow
\beq
\label{habble}
T=\sqrt{\frac{m_pH}{1.66g_*^{1/2}}}=\frac{\sqrt{m_pm_{\theta}}}
{g_*^{1/4}}
\eeq
The last part of expression (\ref{habble}) takes into account that the 
relaxation starts under the condition $H\approx m_{\theta}$.
Using the formulas (\ref{bar1}), (\ref{entropy}), (\ref{habble}) we are able to 
get the baryon/antibaryon asymmetry
\beq
\label{asym}
\frac{n_{B(\bar B)}}{s}=\frac{45g^2}{16\pi^4g_*^{1/4}}\left(\frac{f}{m_p}\right)^{3/2}
\frac{f}{\Lambda}Y(\theta_i)
\eeq
The function $Y(\theta)=\theta ^2
\int\limits_{-\theta /2}^{\theta /2}d\omega\frac{\sin^2{\omega}}{\omega^2}$
takes into account the dependence of amplitude of baryon asymmetry and its
sign on the initial phase value in the different space regions during
inflation.

The expression (\ref{asym}) allows us to get the observable baryon asymmetry of 
the Universe as a whole $n_B/s\approx 3\cdot 10^{-10}$. In the model under 
consideration we have supposed initially that $f\ge H\simeq 10^{-6}m_p$. The 
natural value of coupling constant is $g\le 10^{-2}$. We are coming to 
observable baryon 
asymmetry at quite reasonable condition $f/\Lambda\ge 10^5$ (see e.g. 
\cite{freese}).

\section{\label{DISCUSSION} Discussion}
In this paper we have proposed a model for inhomogeneous baryosynthesis on
the base of the spontaneous baryogenesis mechanism \cite{cohen}. The model
predicts the generation of {\it antizillahs} with sizes exceeding the
critical surviving size. The antibaryon number density relative to background 
baryon 
density 
in the resulting antizillahs and its number depends on the value of phase 
established 
at 
the beginning  and on the parameters of PNG field potential. It is possible to 
have one or several antizillahs the volume box corresponding to every
galaxy depending on the parameter values. The observational consequences
of  existence of antizillahs and the restrictions on their number and
sizes have  been 
analyzed in papers \cite{we,khlop,ams} 

Of course we may in general expect that some region with size exceeding
$l_c$ would contain antibaryon excess after the annihilation of
small primordial domains and antidomains contained in this region is
completed. However the probability to have such region is suppressed
exponentially. Therefore to have observational acceptable number of
antimatter regions \cite{khlop} with the size exceeding the  critical
survival size, a superluminous cosmological
expansion in the formation of primordial antimatter proto--domain seems 
necessary.

As we have mentioned, the additional problem for the most 
models of inhomogeneous baryogenesis invoking phase transitions at the inflation  
epoch is prediction of the large scale unwanted topological defects.  
Our scheme contains the premise for existence of domain walls too. Such
walls are not formed when the only minimum of PNG potential exists, what
corresponds in the considered model to the fluctuations around $\theta
=0$, when the North pole ($\theta =\pi$) is not crossed. But in the case,
when such crossing takes place the multiple degeneracy of vacua appears
(e.g. vacua with $\theta =0$ and $\theta =2\pi$). The equation of
motion that correspondes to potential (\ref{tilt}) admits kink--like,
domain wall solution, which interpolates between two adjacent vacua. 
Thus, when the PNG tilt is significant, domain wall is formed along the
closed surface (e.g. $\theta =\pi$) \cite{kim}. In the other words every
antizillah  with high   relative antibaryon density will be encompassed by
domain wall bag. The wall stress energy $\Delta\approx 8f\Lambda^2$
\cite{kim,sik} leads to the oscillation of wall bag after the whole bag 
enters the cosmological horizon. During the oscillations the   energy
stored in the walls is released in the form of quanta of NG field and  
gravitational waves. As we are taken $0<\theta_{60}<\pi$, the wall's
bag will have the scale of the order of modern horizon, if the  
dispersion $\sigma_{l_{max}}$ is large as $\pi -\theta_{60}$. Owing to
very  large  oscillation period such big wall bag would presumably
survive to the present  time,  
which would be cosmological disaster  
\cite{lyth,lythstewart}. Thus the upper limit on the dispersion will be
$\sigma_{60}<\pi$. From the other hand this condition should be valued if we 
want to have parameters of antizillah population that do not contradict to 
direct and indirect observational constraints \cite{1}. It means that we will 
have wall bags with the sizes less then cosmological horizon and that walls had 
to decay until present time.  The mechanisms of their decay is a subject of 
separate paper, in 
which we also
plan to obtain additional constraints on the model, which follow from the 
condition that walls 
do not dominate within the cosmological horizon before the bag decays. 
If the energy density of walls is sufficiently high
to give local wall dominance in the border
region before the bag enters the horizon,
it is of interest to analyze the role of
superluminous expansion in the border regions
in the bag evolution (see e.g. \cite{vil}). The interesting question on the wall 
interaction with 
baryons in 
the course of wall contraction and decay will be also studied separately.

In general all baryogenesis models that are able to generate some amount of
antimatter regions look like radical limit of models with local baryon number 
density fluctuations so called isocurvature fluctuations 
\cite{LythLiddlePhysRep,iso}. It is known that the contribution of isocurvature 
fluctuations to the cosmic microwave background (CMB) anisotropy obeys to 
$\frac{\delta T}{T}=-\frac{1}{3}\frac{\Omega_B}{\Omega_0}\delta_{B_i}$, where    
$\delta_{B_i}$ is the amplitude of initial baryon number fluctuations and 
${\Omega_0}$ (${\Omega_B}$) are the total (baryon) density (in units of critical 
density). 
As it follows from our numerical illustration (see \ref{PERT} and expression 
(\ref{bar1})) 
we must have quite large amplitude of initial baryon number fluctuations 
$\delta_{B_i}\sim h/\theta_{60}\simeq 10^{-2}$ at the biggest cosmological 
scales, 
and consequently we would have large amplitude of isocurvature fluctuations at
large scales that contradicts with COBE measurements \cite{iso}. 

To be keeping away of the problem of large--scale isocurvature fluctuations, we 
can, for example, prevent the fluctuations of phase at largest cosmological 
scales. The point is that to have antizillah with size exceeding few kpc. we do 
not need to start phase fluctuations at the e--folds that correspond to the 
biggest cosmological scales. It is sufficiently to start fluctuations of phase 
from the moment, for instance, when the scale $8h^{-1}Mpc$ leaves Habble horizon 
during inflation, namely after the $6.2$ e--folds from the beginning of 
inflation. We took this scale, because it is known that at the scale less then 
$8h^{-1}Mpc$ we could be generated initial baryon number fluctuations at the 
level $\delta_{B_i}\simeq 10^{-2}\div 10^{-3}$ without any contradictions with 
observations.          

One of the natural way to prevent the phase fluctuations at the early inflation 
is to keep $U(1)$ symmetry restored during first $7$ e--folds. The mechanism 
that is able to restore symmetry during inflation has been consider in the works 
\cite{lindekofm,lindeax,lindebook,sakhlop}. According to that works we can
introduce 
interaction between inflaton field $\phi$ and field $\chi$. The simple potential 
of such kind may be chosen as
$
V(\phi ,\chi )= \frac{1}{4}\lambda_{\phi}\phi^4+V(\chi )+\nu\phi^2\chi^*\chi
$, where $\nu =m_{\chi}^2/cM_p^2$, and $c\simeq 1$.
The effective mass of the field $\chi$ depends on $\phi$ 
directly $m_{\chi}^2(\phi )= m_{\chi}^2+\nu\phi^2$. One considers here for 
simplicity the case $\nu =m_{\chi}^2/cM_p^2$. This implies that the effective 
value of
mass
$m_{\chi}^2(\phi )$ during inflation is given by $\nu (\phi^2-cM_p^2)$ 
and is positive because of very large value of the inflation field. 
It means that our $U(1)$ symmetry is restored during the period when the 
amplitude of the inflaton field exceeds $\phi_c=\sqrt{c}M_P$, and the field 
$\chi$ settles into the minimum of its symmetric potential.  During this period 
there was no NG boson valley and phase fluctuations. After the moment that 
inflaton field turns to be less then $\phi_c$ the symmetry breaking takes place 
and the NG potential has the radius 
$f_{eff}=\sqrt{\nu(cM_p^2-\phi^2)/\lambda_{\chi}}$ and fluctuations are started. 
To keep symmetry restored during first $7$ e--folds we should have 
$\phi_c=4M_p$. After the moment of symmetry breaking it is allowed to start the 
fluctuations of phase with appropriate dispersion to create antizillahs, without 
any contradictions with observed CMB anisotropy. Of course to evaluate the 
distribution of antizillahs by sizes we have to take another parameters then we
have used in our numerical example, but it does not change the main result of 
this paper.

Another story will take place if we would like to consider the AD
baryogenesis as a basis for generation of antizillahs. 

As it was discussed in the introduction the dynamics of the AD field is
more
complicated that in the case of spontaneous baryogenesis. Moreover it
depends on the fact, D-- or F-- term inflation takes place.
Also some details depend on the dimension ($d=4,6..$) of
non--renormalizable term lifting the flat direction \cite{b3,lisa}, but it
is enough for the brief discussion to circumscribe ourself with
the minimal AD baryogenesis \cite{b3}, where $d=4$. Thus in the case of
D-- term inflation, when the coherent slow rolling of AD field and inflaton
are already established, the maximal radius $f_{eff}^{AD(D)}\simeq
10^{16}GeV$ of effectively massless angular direction can be obtained from
the requirement that radial de Sitter fluctuations of AD field would not
disturb significantly the spectral index of primordial adiabatic density
perturbations \cite{b3} measured by COBE. Thereby, it is possible to get
dispersion of phase fluctuations at the level $h\simeq 10^{-2}$ that is
required for successful generation of antizillahs. The similar situation
we could have in the case of F-- term inflation \cite{b3,lisa} because the
AD potential gets an order of $H^2$ negative mass squared term during
inflation, which causes the effective minimum at
$f_{eff}^{AD(F)}\simeq C_F\sqrt{Hm_p}\simeq 10^{16}GeV$ (the $C_F$ is a
constant of order of one).

The isocurvature fluctuations in the model of inhomogeneous AD
baryogenesis with dispersion of phase fluctuations appropriate for
antizillahs generation should be already observed by COBE \cite{b3}.
Moreover this fluctuations can get some amplification owing to possible
transformation of fluctuations of AD condensate into the isocurvature
fluctuations of neutralinos \cite{b2}. The exact solution of the problem
of isocurvature fluctuations for the AD based antimatter generation is
the subject of separate investigation. Here we can only present some
speculations, how to avoid the large isocurvature fluctuations at large
cosmological scales, which are based on the similar strategy that has been
chosen in the case of spontaneous baryogenesis. 

As it has been
mentioned in the Introduction, to organize the angular effectively massless
direction in the AD potential we should accept the condition of the
absence of order $H$ correction to the A-- term both during and after
inflation \cite{b3}. This condition gets automatically satisfied in the
case of D-- term inflation \cite{mr}, while it is not true if the
inflation is F-- term dominated (see for example \cite{lisa}). According
to this observation we can hope to find the such kind of trajectory of
inflaton in field space that corresponds to the F--
term dominated inflation in the beginning and then goes into D-- term
dominated regime. It implies that during the F-- term dominated inflation
the angular direction gets a mass of order $H$ and imaginary component of
AD field is dumped and exponentially close to the minimum caused by
this effective mass term. In such situation there are no de Sitter
fluctuations of the phase. The fluctuations start only at the moment when
the inflation goes to the D-- term dominated regime and the angular
direction turns to be effectively massless, because there is no correction
of order $H$ to the A-- term anymore. As we estimated before, to put the
maximal scale of isocurvature fluctuations far below the modern
cosmological horizon the transition from F-- term to D--
term inflation should take place 5--10 e--folds after the beginning of
inflation. How to organize such transition is the subject of
separate publication, but it seems that it could appear, for example, in
the context of a realistic supergravity theory deriven from the weak
coupled supestring \cite{string}, which is already beyond the MSSM. There
is some possibility to generate the F-- term from a Fayet--Iliopoulos D--
term \cite{d-f}. It could preserve the flatness of F-- term direction
during the first 5--10 e--folds of inflation causing the F-- term
domination firstly and subsequent trasformation of the vacuum energy into
the D-- term domination mode when it is allowed to begin phase
fluctuations of AD field with dispersion appropriate for generation of
antizillahs and without contradictions with COBE measurements. 

We would like to notice in
conclusion that the regions with antimatter in  matter--dominated Universe
could arise naturally in the variety of models. The main  issue, that is
needed, is a  valley of potential. It is the valleys that are responsible
for formation of causally separated regions with different values of field
which in its turn give rise to antimatter domains. Many extensions of
standard model based on supersymmetry possess this property, what strongly
extends the physical basis for cosmic antimatter searches.

\bigskip
\section{Acknowledgments}
This work was partially performed in the framework of Section "Cosmoparticle 
physics" of Russian State Scientific Technological Program "Astronomy. 
Fundamental Space Research", International project Astrodamus,
Cosmion--ETHZ and  AMS--EPCOS. MYuK and ASS acknowledge supporte from
Khalatnikov--Starobinsky school (grant 00--15--96699). We thank
R.Konoplich and A.Sudarikov for interesting discussions  and
suggestions. We are also grateful to J.Ulbricht, A.D.Linde and I.Tkachev
for useful comments.

\end{document}